# THE FABRE PROJECT AT TRIESTE

G. D'Auria, C. Rossi - Sincrotrone Trieste.
M. Danailov - Laboratorio Fibre Ottiche-Sincrotrone Trieste.
M. Ferrario - INFN Laboratori Nazionali Frascati.
N. Piovella, L. Serafini - Universita' di Milano and INFN.

*Abstract*

A program to design a high brilliance electron source suitable for a short wavelength Linac-based FEL is presented. The goal of the project is to develop a multi-cell integrated photoinjector capable of delivering 1 nC bunches with emittance below 1 mm mrad.

This will be the first step toward a possible development of a IV generation light source test facility based on the existing Trieste Linac. For this purpose a common program between Sincrotrone Trieste and INFN-Milano has been undertaken. Here a brief description of the program and the first results of the RF Gun electromagnetic structure with the beam dynamics on the ELETTRA Linac are presented.

## 1 INTRODUCTION

The growth in interest over the last few years in IV generation light sources based on FELs and the SASE process [1,2] requires further considerable efforts for the production of intense low emittance beams. In fact, in order to reach and operate these facilities at shorter wavelengths, the electron beam emittance $\varepsilon$, and the coherent radiation wavelength $\lambda$, must be very close to each other, to guarantee the maximum overlap between the two beams in phase space. Moreover the efficiency of the SASE-FEL emission process, defined as the ratio between the photon beam power and the electron beam power, scales like $I_{PK}^{1/3}$ with $I_{PK}$ the electron beam peak current. The previous requirements clearly show that the real figure of merit for the electron beam is the normalized brightness, defined as $B_n=I_{PK}/4\pi\varepsilon_n^2$: a FEL in the X-ray band will require $10^{14}$ to $10^{15}$ A/m$^2$ and is nowadays the main challenge to meet. In order to reach these values, that are at the limit of expected performances for the next decade, all the laboratories pursuing a long term program in IV generation radiation sources have begun developing experiments and test facilities to study the physics of high brightness electron beams. The FaBrE project can be viewed within the same context: its main aim is the study and construction of a high brightness photo-injector to be installed on the ELETTRA Linac for a FEL-SASE test facility in Trieste. This ambitious program will also take advantage of the future planned upgrading of ELETTRA: after the commissioning of the new ELETTRA full energy injection system [3], the present 1.0 GeV injector Linac will be available for the proposed test facility.

## 2 AIMS AND TIME SCHEDULE

Despite the large progress in brightness seen in the 90's by electron sources based on RF photo-injectors, only recently has a theoretical understanding of the phenomenon of emittance degradation, hence of the achievable brightness by a photo-injector, been made possible. The success of the "invariant envelope" model, that concerns the prediction of a new equilibrium mode for a beam in the laminar flow regime, [4,5], will allow further progress in this strategic field enabling an optimum control of the emittance growth.

On this basis it has been shown recently that the "integrated" photo-injector, whose first accelerating section is integrated into the gun itself, has the optimal configuration to produce a beam satisfying the invariant envelope conditions when compared to the more popular "split" version, where the accelerating section is physically separated from the gun.

Starting from these considerations the goal of the FaBrE project is the study and the construction of an integrated photo-injector whose RF structure complies, as much as possible, with the requirements imposed by theory:

i) high spectral purity of the accelerating field profile on the axis;
ii) normalized amplitude of the accelerating field, $\alpha=eE_0/2k_{RF}m_ec^2=1.3$, corresponding at S-band to a peak field of 80 MV/m;
iii) shunt impedance as high as possible (compatible with the first requirement);
iv) capability to host a photo-cathode in ultra high vacuum ($\leq 10^{-9}$ torr);
v) stability and large separation between contiguous resonant modes in the operating band.

In order to distribute as much as possible costs and commitments, the whole program will be divided into three different phases.

**Phase I**, with a quite modest commitment and cost, will be limited to theoretical studies for the choice of the most suitable electromagnetic structure, and the construction of a whole accelerating section with its electro-magnetic characterization at low RF power. On the basis of the first numerical simulations carried out on both Coupled Cavity Linac (CCL) and Plane Wave Transformer (PWT) accelerating structures, it seems that the second solution has a better matching with the theoretical requirements previously mentioned [6], even if more simulations are required. We plan to get the first copper model of the accelerating section before the end of this year. This phase, already funded, should be completed at the beginning of 2001 with the RF characterization of the accelerating structure.

**Phase II**, in which the first prototype of the final accelerating structure will be assembled and tested at high RF power levels. The same section will also be used for preliminary beam tests using a commercial Q-switched Nd:YAG laser (Quantel YG 585_10), not synchronized with RF, but already available in our laboratory. Preliminary estimations show that with a slight modification of the laser cavity and utilizing a BBO crystal for IV harmonic generation, this laser can deliver 7 nsec pulses with 10 mJ at 266 nm in a nearly gaussian transversal mode. An additional shortening and smoothing of the delivered pulses are also expected by further optimization of the laser cavity and, if necessary, using a slicing with fast Pockels cell (the latter may turn out to be necessary in order to avoid cathode damage). The goal of the second phase, not yet funded, is to have within 2002 a 12 to 20 MeV electron beam, with 100 to 200 A peak current, and less than 5 mm mrad normalized emittance.

**Phase III**, the photo-injector will be upgraded with a mode-locked laser system able to deliver very short UV pulses (230 nm, 10 ps, 200 µJ, with rise time and jitter less than 1 ps). Such a performance can nowadays be obtained by using several different types of mode-locked systems (i.e. based on Nd:glass, Cr:LiSAF or Ti:Sapphire); the final choice will be made on the basis of further numerical simulations as well as on the experience gained from phase II. At the end of this phase we expect to have a 20 MeV pulsed electron beam with roughly 150 A peak current, 1 nC charge, 1 mm mrad emittance and energy spread better than 1%. The expected quality of the electron beam will allow the start-up of the suggested R&D program installing the photoinjector on the existing Linac.

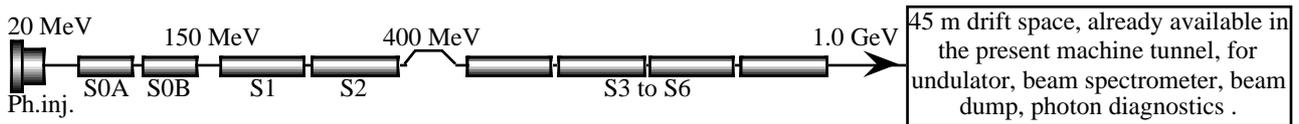

Figure 1: Machine layout

# 3 PRELIMINARY BEAM DYNAMICS SIMULATIONS OF THE ELETTRA LINAC

## 3.1 General Layout of the machine

A complete desciption of the Trieste Linac can be found in [7,8]. Neglecting the electron source and the bunching section, the whole machine consists of two different parts: i) a 100 MeV preinjector, made up of two 3.2 m long constant impedance accelerating sections (S0A, S0B), equipped with focusing solenoids (up to 2.5 KGauss) and presently operated at 18 MV/m; ii) the second part of the Linac is made up of seven 6.2 m long BTW accelerating sections (S1 to S7) equipped with a SLED pulse compressor system. Eight Thomson TH2132 45 MW klystrons feed the whole machine. The maximum operating gradient has been reached at 28 MV/m, and the machine can provide a maximum energy of 1.2 GeV. Keeping fixed the first 100 MeV supplied by the preinjector the remaining sections on average can easily supply 150 (or 100) MeV/section with (or without) SLED respectively. However one of the 6.2 m sections will be used for the new full energy injector and the operational energy of the test facility will be between 0.7 and 1.0 GeV (with/without SLED).

## 3.2 Preliminary beam dynamics studies

Preliminary beam dynamics studies have been carried out on the whole ELETTRA Linac to estimate the beam parameters that could be obtained after installation of the photoinjector (following phase III). All of the simulations have been made using the semi-analytical code HOMDYN and considering the emittance compensation theory. Even if in the future, to reach lower emittance, different machine layouts could be considered, at present, in order to minimize costs, we have implemented only a slight modification of the present machine layout, shifting the third Linac section, S3, and leaving a 10 m drift space between sections S2 and S3 to install the magnetic bunch compressor, see Fig. 1 for a simplified scheme of the machine layout. Magnetic optics between the sections is not considered in these preliminary studies.

The photoinjector parameters have been optimized in order to get a laminar waist and a maximum of the relative emittance at the entrance of S0A, that results to be a suitable condition to damp the emittance oscillations [9]. The gun energy has been fixed at 20 MeV and the matched accelerating field of the first two structures results to be 21 MV/m exiting the second structure at 150 MeV, see Fig. 2 and 3.

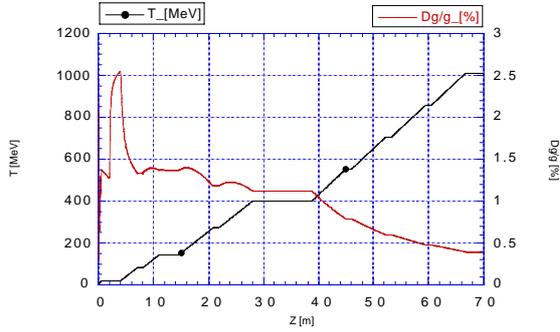

Figure 2: Energy gain and energy spread along the Linac.

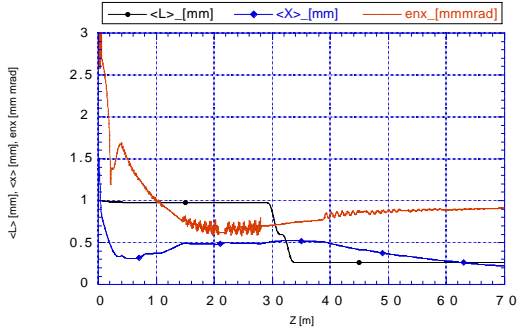

Figure 3: Bunch length, beam envelope and transverse emittance along the Linac.

As shown in Fig. 3, in the drift downstream of S0B the emittance approaches its absolute minimum. The beam is then injected in the first two sections of the second Linac, S1 and S2, 20 degrees off crest at 21 MV/m to provide the necessary energy spread for the magnetic compression expected at 400 MeV. The magnetic compressor is modeled by HOMDYN as a "one wiggler period" according to the wiggler hard edge model reported in [10]. The equivalent period results to be $\lambda_w=\pi^2 l_{bend}$ where $l_{bend}$ is the dipole magnet length and the equivalent field strength is $B_w=4B_{bend}/\pi$. In the present design $l_{bend}$=0.5 m and $B_w$=0.11 T; a focusing gradient in the chicane has also been considered with a suitable pole shaping. At the exit of the compressor the average peak current is 300 A, but inside the bunch, as shown in Fig. 4, the slice peak current reaches higher values over approximately one quarter of the bunch length.

In the remaining 4 structures the beam is driven up to 1 GeV with a further energy spread reduction to 0.4 %. The slight emittance growth shown in Fig. 3 after the beam compression, is mainly due to a lack in the required optics to compensate for space charge effects induced by bunch compression.

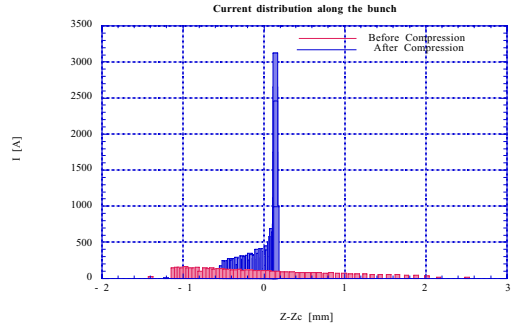

Figure 4: Current distribution along the bunch before (lower curve) and after the magnetic compression.

## 4 CONCLUSIONS

An ambitious program for a high brilliance electron source based on a 20 MeV integrated photoinjector has been recently initiated at Sincrotrone Trieste under the FaBrE collaboration. The preliminary results are encouraging and in the near future the beam dynamics simulations will be extended to include the Linac optics and the effects of the beam interaction with the undulator.

## REFERENCES

[1] R. Bonifacio et al., "Collective instabilities and high-gain regime in a Free Electron Laser", *Opt. Commun.* 50, 1984.
[2] J. Arthur et al., "Linac Coherent Light Source (LCLS) Design Study Report", SLAC-R-521, April 1988.
[3] C.J. Bocchetta et al., "A full energy injector for Elettra", EPAC 2000, Vienna, June 2000.
[4] L. Serafini et al., "Envelope analysis of intense relativistic quasilaminar beams in RF photo-injector: a theory of emittance compensation", Phys. Rev. E, Vol. 55, June 1997.
[5] L. Serafini et al., "New generation issues in the beam physics of RF laser driven electron photo-injectors", SPIE-LASER '99 Conf., San Jose', CA, January 1999.
[6] G. D'Auria et al., "The FaBrE project: design and construction of an integrated photo-injector for bright electron beam production", EPAC 2000, Vienna, June 2000.
[7] D. Tronc et al. "The ELETTRA 1.5 GeV electron injector", PAC '91, S. Francisco, May 1991.
[8] G. D'Auria et al., "Operation and status of the ELETTRA injection Linac", PAC '97, Vancouver, May 1997.
[9] M. Ferrario et al., "HOMDYN study for the LCLS RF photo-injector", LNF-00/004 (P), SLAC-Pub 9400, March 2000.
[10] H. Wiedemann, "Particle Accelerator Physics", Spring-Verlag, 1993.